\documentclass[aps,pra,twocolumn,amsmath,amssymb,nofootinbib,superscriptaddress]{revtex4}
\usepackage{graphicx}
\usepackage{epstopdf}
\newcommand{\bra}[1]{\langle#1|}
\newcommand{\ket}[1]{|#1\rangle}
\newcommand{\vac}{\mathrm{vac}}
\newcommand{\eme}{\mathrm{EME}}

\newcommand{\eqn}[1]{Eq.~(\ref{#1})}
\newcommand{\fig}[1]{Fig.~\ref{#1}}

\begin{document}
\title{Scalable quantum computing with atomic ensembles}

\author{Sean D. Barrett} \email[]{seandbarrett@gmail.com}
\affiliation{Blackett Laboratory, Imperial College London, Prince Consort Road, London SW7 2BZ, United Kingdom}

\author{Peter P. Rohde} \email[]{peter.rohde@materials.ox.ac.uk}
\affiliation{Department of Materials, University of Oxford, Parks Road, Oxford, OX1 3PH, UK}

\author{Thomas M. Stace} \email[]{stace@physics.uq.edu.au}
\affiliation{Department of Physics, University of Queensland, Brisbane, QLD 4072, Australia}

\date{\today}

\frenchspacing


\begin{abstract}


Atomic ensembles, comprising clouds of atoms addressed by laser
fields, provide an attractive system for both the storage of quantum
information, and the coherent conversion of quantum information between
atomic and optical degrees of freedom. In a landmark paper, Duan et
al. (DLCZ) \cite{bib:Duan01} showed that atomic ensembles could be
used as nodes of a quantum repeater network capable of sharing
pairwise quantum entanglement between systems separated by
arbitrarily large distances. In recent years, a number of promising
experiments have demonstrated key aspects of this proposal
\cite{PhysRevLett.93.233602,balic:183601,lan:123602,Chen08,Choi08, Simon07}. Here, we describe a scheme for full
scale quantum computing with atomic ensembles. Our scheme uses
similar methods to those already demonstrated experimentally, and
yet has information processing capabilities far beyond those of a
quantum repeater.

\end{abstract}

\maketitle

Amongst the more promising schemes for the implementation of
scalable quantum computing are those in which qubits are stored in
individual trapped atoms and entangled via single photon
interference of photons emitted by the atoms \cite{bib:BarrettKok05,
bib:Lim05}. An appealing aspect of such schemes is that once
high-fidelity elementary one- and two-qubit operations can be
demonstrated, it is in principle straightforward to scale to a large
number of qubits. Notwithstanding recent experimental progress
\cite{Moehring07}, this approach remains
experimentally challenging due to the difficulty of trapping and
manipulating single atoms, and  the difficulty of coupling a single
atom to a single optical cavity mode to improve photon collection
efficiency. 

Atomic ensembles provide a promising alternative system.
The atomic excitations of a cloud of $N$ identical atoms are
strongly coupled to the optical field through \emph{collective
enhancement} \cite{bib:Duan01}, which increases the effective
atom-photon coupling by a factor of $\sqrt{N}$ over the single atom
case, negating the requirement for a cavity. Furthermore, it is not
necessary for the atoms to be in the Lamb-Dicke limit
\cite{duan2002tdt}, significantly reducing the trapping and cooling
requirements.

\begin{figure}
\begin{center}
\includegraphics[width=\columnwidth]{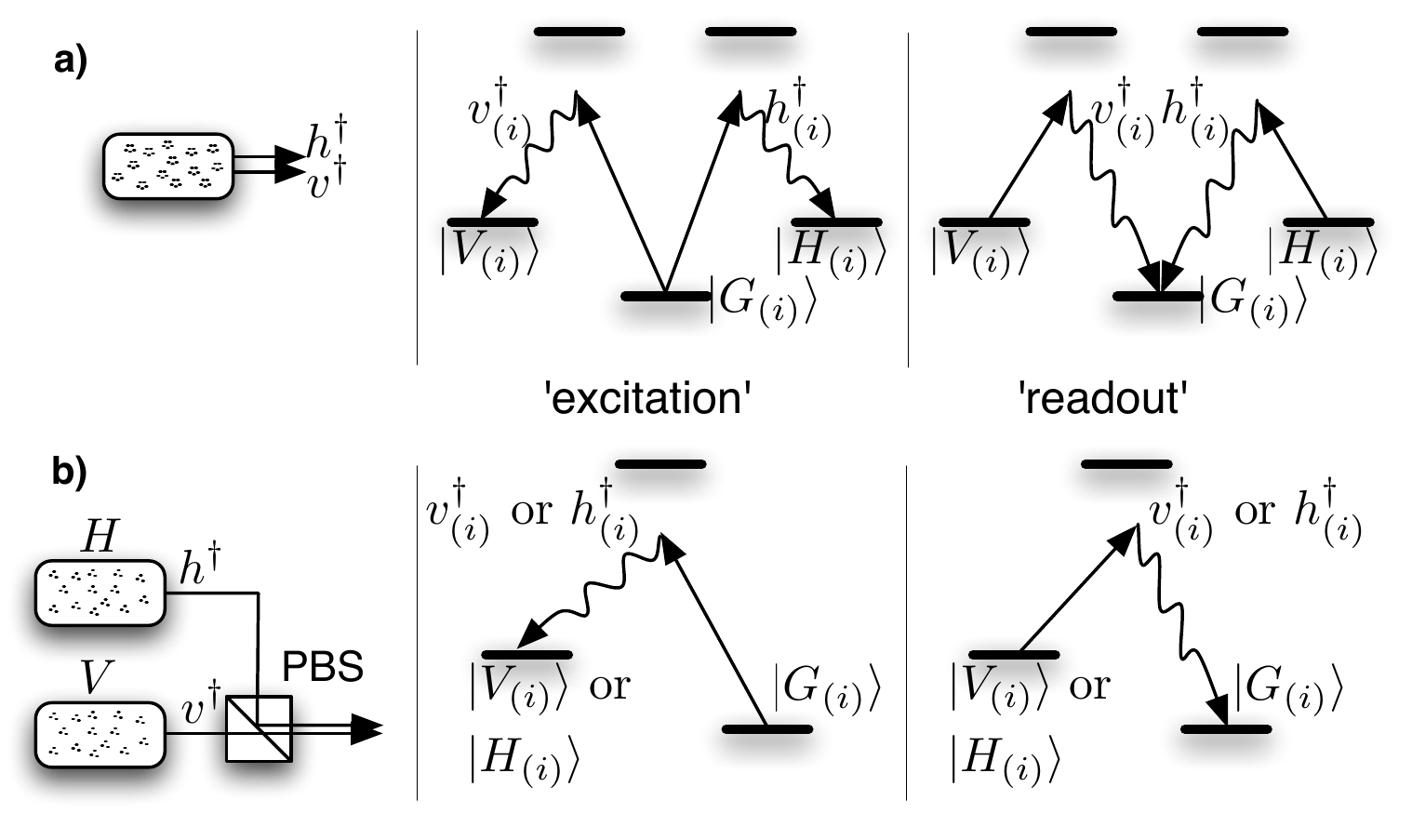}
\caption{Configuration and atomic level structure for (a)
internal-state encoding: a qubit is encoded in the two internal
atomic states $\ket{H_{(i)}}$ and $\ket{V_{(i)}}$, of atoms in a
single ensemble and (b) dual-rail encoding: a qubit is encoded in
the single atomic state of atoms in two separate but identical
ensembles, labelled `$H$' and `$V$'. The ensembles may be two
distinct regions within the same vapour cell, addressed by spatially
separated lasers. When operated in `excitation' mode, a weak, off
resonant laser field drives the upward transition (straight lines),
and a Stokes photon is emitted (wiggly lines).  In `readout' mode,
the frequencies are reversed, producing an anti-Stokes photon in the
output mode.} \label{Levels}
\end{center}
\end{figure}

We encode a logical qubit in the collective excitations as
$\ket{0}_L\equiv\ket{H} = H^\dag\ket{G}$ and $\ket{1}_L\equiv\ket{V}
= V^\dag\ket{G}$. Here $S^\dag=N^{-\frac{1}{2}}\sum S^\dag_{(i)}$
(for $S^\dag=H^\dag$ or $V^\dag$) represent symmetric collective
excitations of the ensemble, where $H_{(i)}^\dag = \ket{H_{(i)}}
\bra{G_{(i)}}$ denote the excitation operator for atoms, and
$\ket{G}=\ket{G_{(1)}G_{(2)}...G_{(N)}}$ is the state with all atoms
in the ground state. Two alternative encodings are possible: an
internal-state encoding, where the qubit is encoded in two internal
atomic levels, in which case $H_{(i)}/V_{(i)}$ refer to the two
metastable energy levels in \fig{Levels}a; alternatively, a dual
rail encoding, where the qubit is encoded in a pair of identical
ensembles, each consisting of atoms with the simpler level structure
of \fig{Levels}b, in which case $H/V$ label the relevant ensemble.
Note that the two ensembles in \fig{Levels}b may simply be two
spatially distinct regions within the same atomic vapour cell,
addressed by independent lasers.

A collective excitation, together with a forward-scattered Stokes
photon (`excitation' panels of Fig. \ref{Levels}), can be generated
by weakly driving one arm of the corresponding Raman transition,
yielding the state
\begin{equation}
\ket{\psi} = \ket{G}\ket{\vac} + \sqrt{p} S^\dag s^\dag \ket{G}\ket{\vac}
+ O(p),
\end{equation}
where $\ket{\vac}$ denotes the vacuum state of the optical mode,
$s^\dag$ the corresponding photon creation operation operator (which
could be $s^\dag=h^\dag$ or $v^\dag$), and $p \ll 1$ is the
probability of exciting the ensemble. The third term, corresponding
to multiple excitations, can be made arbitrarily small relative to
the second term by reducing the excitation probability $p$.  The
collective excitation may be measured destructively during
`readout', shown in Fig. \ref{Levels}, by driving the reverse
transition, resulting in the conversion of the atomic excitations
into anti-Stokes photons; the efficiency of this process can be
close to unity owing to collective enhancement
\cite{duan2002tdt,Simon07}.

Since the qubits are each encoded in separate ensembles (or ensemble
pairs) there is no direct interaction between the qubits. Our
proposal thus follows DLCZ \cite{bib:Duan01} and entangles separated
ensembles using linear optics networks and photodetection on the
coherently scattered light. Such entangling operations are
non-deterministic, with success heralded by a particular sequence of
photodetector clicks. To overcome this indeterminism, we use a form
of measurement based QC \cite{bib:Raussendorf01}, in which heralded
entangling operations can be used to efficiently construct an
entangled \emph{cluster state} of many qubits \cite{bib:Nielsen04,
bib:BrowneRudolph05, bib:BarrettKok05, bib:LimBarrett05}. Such
states are described by a graph comprising a collection of edges
between qubits. The cluster state corresponding to this graph is
defined as the state that results from initializing each qubit in
the state $\ket{+} = (\ket{0}+\ket{1})/\sqrt{2}$, and then applying
controlled-phase operations ${\mathrm{CZ}_{ij} =
\mathrm{diag}(1,1,1,-1)}$ to all qubit-pairs $i$ and $j$ linked by
graph edges. Once the cluster state has been prepared, universal QC
can be implemented by a sequence of single qubit measurements on the
state of the form $\sin(\theta_i)X_i + \cos(\theta_i)Y_i$. Here
$X_i$, $Y_i$, $Z_i$ are the Pauli operators on the $i$th qubit, and
$\theta_i$ is a parameter that depends on the outcomes of earlier
measurements. These measurements are implemented by mapping the
ensemble qubit onto a photon (`readout' of \fig{Levels}),
transforming the photon polarization using standard linear optical
elements, then measuring the photon in the $h$-$v$ basis.

In the remainder of this paper, we describe a procedure for building
a cluster state of atomic ensemble qubits using linear optical
networks and photodetection. We first outline a process for
\emph{fusing} two arbitrary cluster states together via an optical
network that implements a heralded controlled-phase operation. We
then describe a protocol for constructing primitive 3-qubit cluster
states. Cluster fusion plus 3-qubit cluster states is then
sufficient to build arbitrary clusters. The protocol is first
described for the idealized case of  perfect `readout', perfect
collection and detection efficiency, negligible unwanted
excitations, and negligible decoherence of the atomic ensemble
qubits. We subsequently argue that the scheme is also robust when
these idealizations are relaxed.

For clarity of exposition, we assume that arbitrary local unitary
mode transformations can be performed on each ensemble (or pair of
ensembles) representing a single qubit. In fact, such operations may
be difficult to implement in practice; however, as we describe in
the supplementary information, all such operations can be deferred
until after the atomic excitations have been mapped onto photon
states, and thus can be implemented with standard linear optical
elements.

\begin{figure}
\begin{center}
\includegraphics[width=\columnwidth]{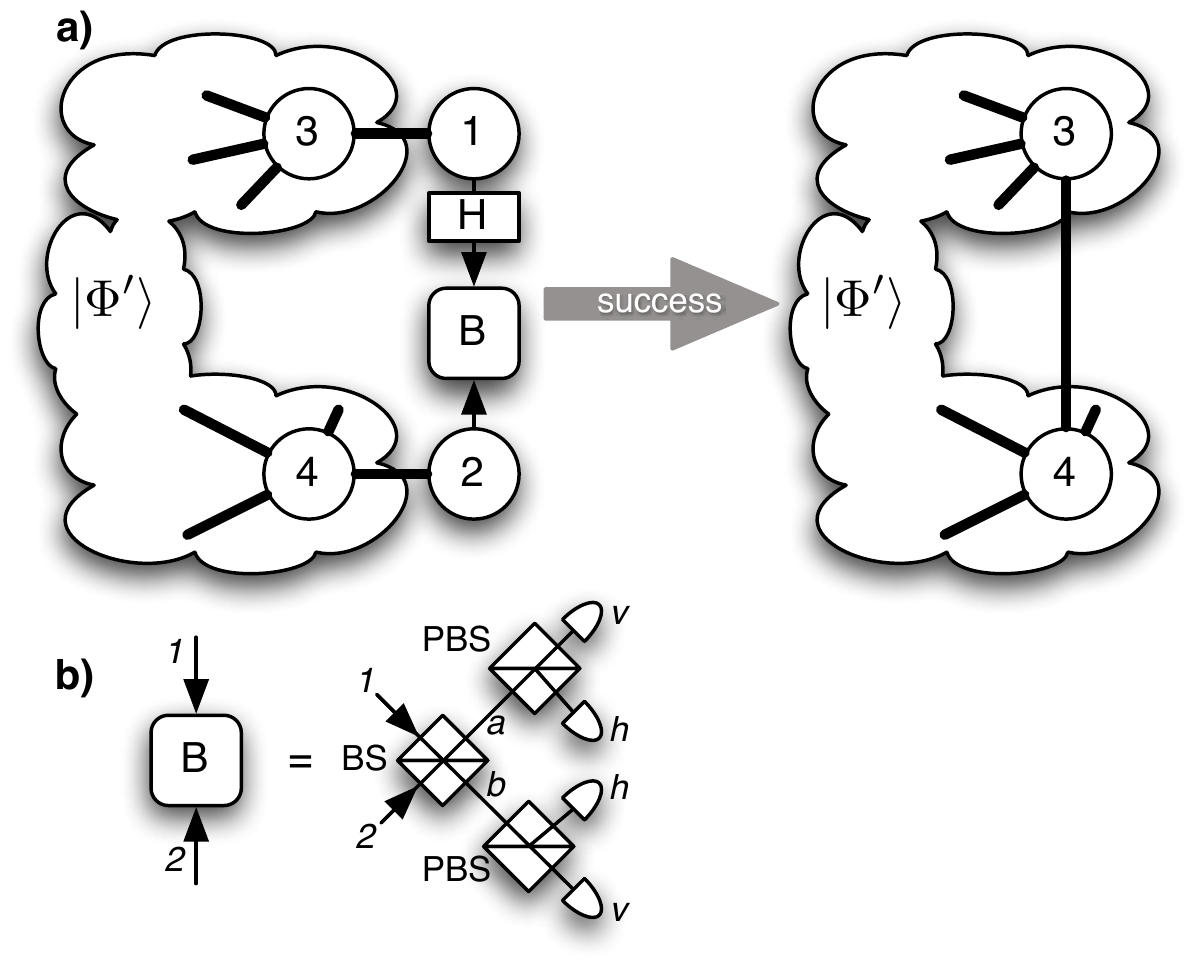}
\caption{a) Protocol for implementing a destructive CZ gate between
nodes of a cluster state, each designed to have a singly-linked node
(nodes 1 and 2), with which to build the link.  Readout pulses are
applied to ensembles 1 and 2, converting the atomic excitations into
photons.  The box labelled $\mathsf{H}$  is a Hadamard gate on the
qubit originally encoded in ensemble 1, which can be implemented via
linear optical elements. Note that $\ket{\Phi'}$ may be a product of
two disconnected cluster states, so this operation fuses the two
together.  b) Box $\mathsf{B}$ consists of a beam splitter (BS) and
two polarisation sensitive photodetectors. A successful CZ operation
is implemented between qubits 3 and 4 when the detected photons are
found to have opposite polarisations.} \label{Fusion}
\end{center}
\end{figure}
We now describe how to apply a controlled-phase (CZ) gate between
two qubits, labelled 3 and 4 in \fig{Fusion}a, each attached to an
arbitrary cluster, and each attached to singly linked qubits
(labeled 1 and 2). This gate is \emph{destructive} since it consumes
qubits 1 and 2, and implements a CZ gate between their neighbours.
This operation serves to forge new links between clusters, thereby
building larger or more connected graphs.

The (unnormalized) state in \fig{Fusion}a may be written
\begin{equation}
\ket{\Phi}=(H_1^\dag+V_1^\dag Z_3)(H_2^\dag+V_2^\dag
Z_4)\ket{\Phi'}\ket{G}_1\ket{G}_2.
\end{equation}
The collective excitations in ensembles 1 and 2 are converted to
photons using the readout pulse of \fig{Levels}.  The optical output
of 1 passes through a Hadamard gate, $\mathsf{H}$, then is mixed
with the output of  2 in the optical circuit shown in \fig{Fusion}b.
Immediately before the photons arrive at the photodetectors, the
state of the system is given by
\begin{eqnarray}
\ket{\Psi}&=&[(h_a^\dag v_a^\dag-h_b^\dag v_b^\dag)
Z_4 \textrm{CZ}_{34}\nonumber\\
&&{}-(h_b^\dag v_a^\dag-h_a^\dag v_b^\dag)Z_3Z_4 \textrm{CZ}_{34}\\
&&{}+(({h_a^\dag}^2 - {h_b^\dag}^2)(1+Z_3)\nonumber\\
&&{}+({v_a^\dag}^2 - {v_b^\dag}^2)Z_4
(1-Z_3))]\ket{\Phi'}\ket{\vac}_{ab},\nonumber
\end{eqnarray}
where $a$ and $b$ label the output modes from the beam splitter. The
first two terms in $\ket{\Psi}$ represent states in which the two
photons have different polarisations, and the last term represents
states in which the two photons have identical polarisations. Thus
when the photodetectors click, with 50\% chance the photons are
found to have opposite polarisations, and the state is successfully
projected onto one that is equivalent (up to local $Z_i$ operations
which must be applied conditional on the outcome) to having a CZ
gate applied between ensembles 3 and 4, resulting in a new cluster
state with a link between nodes 3 and 4, as shown in \fig{Fusion}a.
When photons are registered with the same polarisation, the gate
fails, and ensemble 3 is projected onto one of the eigenstates of
$Z_3$, which has the effect of removing it from the cluster. Thus,
in the event of a failure, a total of three ensembles are removed
from the initial state.

A cluster state can be efficiently constructed with this
non-deterministic but heralded CZ operation: failure of any
operation damages the graph, but this can be repaired in subsequent
steps. Provided an appropriate strategy is used, the total cost of
preparing the state is proportional to the size of the cluster
\cite{bib:Nielsen04, bib:BrowneRudolph05, bib:BarrettKok05,
bib:LimBarrett05,bib:RohdeBarrett07,bib:Kieling06,bib:Gross06,kieling:130501}.



\begin{figure}
\begin{center}
\includegraphics[width=0.9\columnwidth]{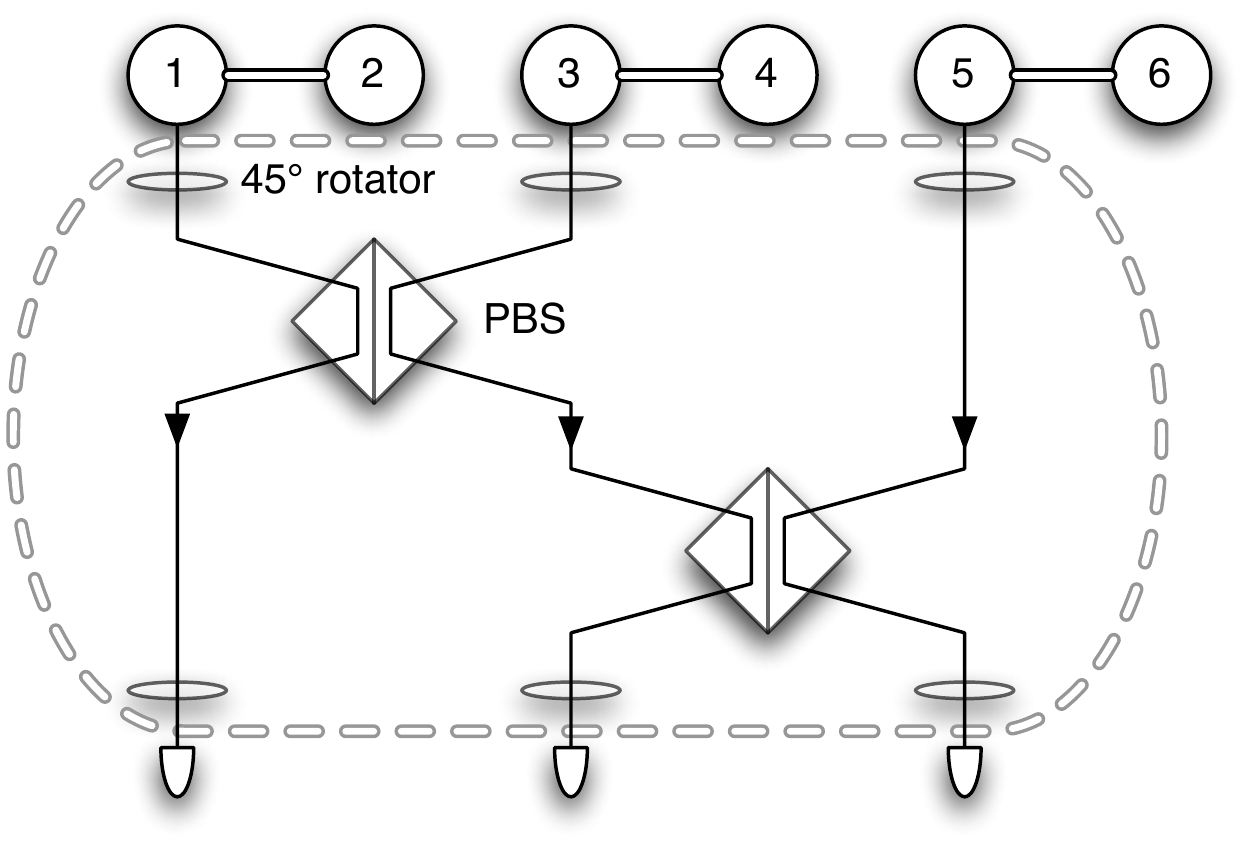}
\caption{Protocol for transforming three two-qubit EME states into a
three qubit cluster state using polarisation rotators and polarising
beam splitters (PBSs). Numbered circles represent atomic ensembles,
double lines indicate EME correlations, and thin lines represent
optical paths.  The dashed line encloses linear optical elements.
Readout pulses are applied to qubits 1, 3 and 5. If each of the
three detectors register a single photon, ensembles 2, 4 and 6 are
projected onto a state that is equivalent, up to local
transformations, to a three-qubit cluster state.} \label{EMEGHZ}
\end{center}
\end{figure}

With our CZ gate, successful fusion of an $n$ qubit cluster and an
$m$ qubit cluster results in a new cluster of size $n+m-2$. Thus, in
order to grow large cluster states, we require an initial supply of
3-qubit cluster states. We now describe a two-step recipe for
preparing these states, starting from an initial supply of ensembles
in the $\ket{G}$ state.

The first step is to prepare three copies of an `effective maximally
entangled' (EME) state of two ensembles, which is given by
$\ket{\eme}_{i,j}=(H_i^\dag+V_j^\dag)(V_i^\dag+H_j^\dag)\ket{G}$
\cite{bib:Duan02}. These states are `effectively' entangled, in the sense that
projecting them onto the subspace with a single excitation per
ensemble results in a maximally entangled state of two qubits. EME
states of two ensembles (or ensemble pairs) may be prepared by first
applying a weak $h^\dag$ excitation pulse to each ensemble, mixing the
forward scattered Stokes photons on a 50/50 beamsplitter, and then
detecting the photons with photodetectors. If a single photon is
registered, the ensembles are left in one of the states $(H_i^\dag
\pm H_j^\dag)\ket{G}$, depending on which detector clicked.
Repeating the procedure with $v^\dag$ excitation pulses results in the
state $(H_i^\dag \pm H_j^\dag)(V_i^\dag \pm V_j^\dag)\ket{G}$. By
applying appropriate local unitary mode transformations, this state
can be brought into the form $\ket{\eme}_{i,j}$.

While these states may be useful in certain small scale applications
\cite{bib:Duan02}, the existence of multiple excitation terms are
problematic for a scalable quantum computing scheme, since they
correspond to \emph{leakage errors} in the computation. We overcome
this problem with the network of \fig{EMEGHZ}. This takes as input
three EME pairs, and conditional on the correct sequence of
measurement outcomes, outputs a \emph{true} three-qubit maximally
entangled cluster state, with only a single excitation per qubit.
This network was inspired by the observation that a similar network,
in an all-optical QC context \cite{bib:Varnava07}, also removes
double excitation terms. Given a supply of ideal EME states, the
success rate for this step is 1/32 (c.f. \cite{bib:Varnava07}).


Our destructive CZ gate together with this initial supply of 3-qubit
clusters is sufficient to build arbitrary cluster states with a
total cost linear in the size of the clusters. To give an estimate
of this cost, our (as yet unoptimized) scheme can produce an
$N$-ensemble linear cluster state with a total of $1536 N / p$
elementary laser operations (see supplementary information). (Note
that $p$ should be made sufficiently small to reduce the rate of
multiple excitation errors in the state).



Physical processes which lead to errors in the computation include
atomic decoherence, losses in linear optical elements, imperfect
coupling between collective atomic and optical modes \cite{Simon07},
and imperfect photo-detection and dark counts
\cite{bib:RohdeRalph06b}.  Fault tolerant quantum computation (FTQC)
architectures exist for non-determinsitic measurement-based schemes
\cite{bib:Dawson05, bib:Dawson06, raussendorf2006fto}, so we do not
address the general question of how to correct \emph{all} errors in
our scheme, but note that FTQC can be implemented, provided the
total error rate lies below the FTQC threshold, which is around
$10^{-3} - 10^{-2}$.

The dominant sources of error in this proposal -- imperfect coupling
efficiency, photon loss and detector inefficiency -- are quantified
by  the effective readout probability $\eta$, that an excitation in
a collective atomic mode is firstly mapped by a `readout' pulse into
the correct optical mode, then propagates through the optical
network and is ultimately detected by a photo-detector.  A readout
failure thus leads to a \emph{heralded loss} error, signified by the
absence of a photo-detection event. In the supplementary information
we draw on previous work \cite{bib:Varnava07} to show that heralded
loss errors can be tolerated provided $\eta>2/3$, which is much less
restrictive than FTQC thresholds for more general error models.

This result is particularly promising for our scheme since many
decoherence  processes such as thermal motion, atomic dephasing and
spontaneous emission simply lower the  coupling rate between
collective atomic and optical modes \cite{bib:Duan01,Simon07}. These
errors simply reduce $\eta$ (rather than producing logical errors),
so can be tolerated with the very modest threshold for heralded loss
errors. This robustness is a consequence of the redundant encoding
of the logical qubits in collective modes of many atoms.

We have described a scalable scheme to perform quantum computation
with atomic ensembles and linear optics. The scheme uses similar
methods to those used in quantum repeater experiments, and yet our
proposal has information processing capabilities far beyond those of
a quantum repeater. An important aspect of the scheme is the
efficient elimination of doubly-excited components in the created
entangled states. A reasonable near future experimental goal is the
creation of  heralded 3 qubit cluster states  (Fig. \ref{EMEGHZ}). This
involves only a moderate increase in complexity over existing
experiments, and would allow interesting applications such as
multi-party tests of quantum mechanics 
\cite{greenberger1990bst}, and quantum
secret sharing  \cite{PhysRevA.59.1829} in a non-postselected setting.

We thank Ian Walmsley, Virginia Lorenz, Josh Nunn, Simon Benjamin, Gerard Milburn,
and Terry Rudolph for useful conversations. SDB is supported by the
EPSRC.  PPR thanks the QIPIRC (No.\ GR/S82176/01) for support.  TMS is supported by the ARC.
%

%

\clearpage




\section{Supplementary material: Overcoming loss errors} \label{proof_GHZ}

In this section we consider in more detail the effect of various
loss processes within our scheme. In particular, the dominant loss
processes are due to photodetector inefficiency (that is, the effect
of photodetectors failing to register a `click' when a photon enters
a detector) and the effect of imperfect coupling between the atomic
ensemble and the correct forward-scattered photon mode. Imperfect
ensemble-photon coupling arises from a number of physical processes.
Firstly, there is a fundamental limit imposed by the competition
between collective coupling of the ensemble to the forward scattered
mode, and single-atom spontaneous emission into other free-space
modes \cite{bib:Duan01,Simon07}. Secondly, thermal motion of the
atoms can reduce the efficiency of the ensemble-photon conversion
process \cite{Simon07}. In addition, a variety of dephasing and
relaxation processes which act on the atoms in the ensembles can
reduce the effective coupling efficiency \cite{Simon07}.

In this work, we model detector inefficiency by replacing each
inefficient detector with a perfect one, preceded by a beamsplitter with transmissivity $\eta_D$. Similarly, imperfect ensemble-photon coupling
processes can be modeled by assuming the ensemble-photon coupling is
perfect, but adding a beamsplitter with transmissivity $\eta_E$ on
the output of each ensemble.

An important technique for analysing the effect of these losses is
to note that, formally, these beamsplitters can be commuted through
other linear optical elements, with the aid of the commutation
relations given in \cite{bib:Varnava07}. To further simplify the analysis
we make the assumption that each detector has the same efficiency
$\eta_D$, and that the ensemble-photon coupling efficiency takes the
same value $\eta_E$ for every ensemble. We also, for the purposes of
this section, ignore all other imperfections such as detector dark
counts. This assumption is justified since with modern APD
detectors, dark count rates are typically rather low ($\sim$ 50
s$^{-1}$).

In the remainder of this section, we consider the effect of these
losses at each stage of our protocol, starting at the lowest level
(creating two-ensemble EME states), and then considering the higher
level processes of creating 3-qubit GHZ states, and bonding clusters
with the destructive CZ gate. Our aim is to show that, to a good
approximation, these errors can lead to an `independently degraded'
(ID) error model, where we can form states that correspond to
initially ideal cluster states which have subsequently been subject to
uncorrelated qubit loss errors acting independently on each qubit.
The effect of such loss errors is to place the affected ensemble
qubit in the $|G\rangle$ state, regardless of it's initial state.
These ID states can then be used to perform FTQC with a very high
threshold \cite{bib:Varnava07}, corresponding to a loss probability of
$0.5$ per qubit. Throughout this section, we make substantial use of
methods and results due to Varnava et al. \cite{bib:Varnava07}, who
discussed the issue of loss tolerance in the context of all-optical
quantum computing.

\subsection{Effect of losses in EME state preparation}
\begin{figure}
\begin{center}
\includegraphics[width=\columnwidth]{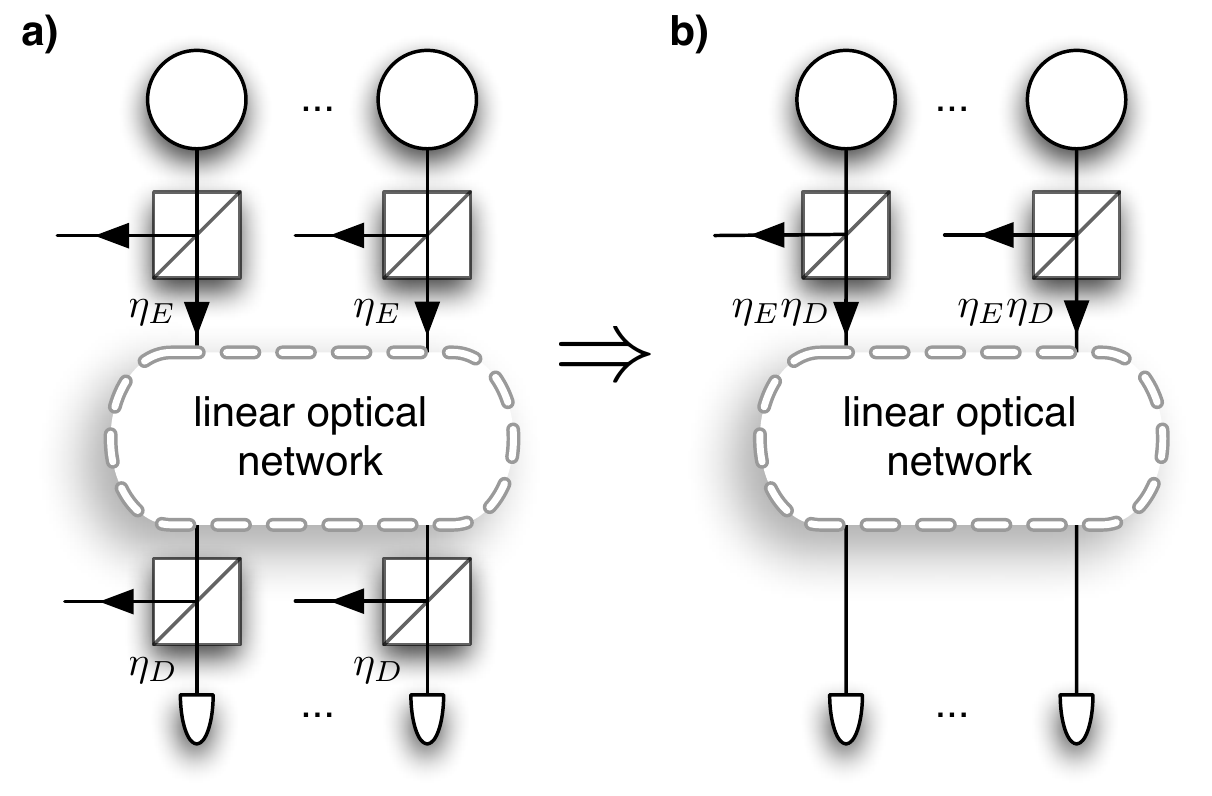}
\caption{Commuting detector loss through the beamsplitter in the EME
preparation circuit so as to map it to source loss. Now the loss in
the system can be characterized in terms of a single efficiency
parameter $\eta = \eta_E \eta_D$ acting on the source.}
\label{fig:EMELoss}
\end{center}
\end{figure}

Here we consider the effect of loss at the lowest level of the
protocol, i.e. when we try and make EME states of two ensemble
qubits. As noted above, there will be two types of loss error at
this stage of the protocol, characterized by the detection and
coupling efficiencies, $\eta_D$ and $\eta_E$, which we can model by
beamsplitters with the corresponding transmissivities (see
\fig{fig:EMELoss}a). The first step in the analysis is to commute
the beamsplitters $\eta_D$ to the output of each ensemble, leading
to an effective model in which the detectors are perfect, but each
ensemble has a beamsplitter of effective transmissivity
$\eta=\eta_D\eta_E$ placed at its output (\fig{fig:EMELoss}b).

Consider a single ensemble, together with the $\eta$-beamsplitter at
its output. Immediately after an excitation pulse (acting on, for
example, the $h$ Raman transition of the ensemble), the combined
state of the ensemble and the optical fields is \cite{bib:Duan01}
\begin{align}
|\psi\rangle_i & = \sum_{n} \frac{p^{\frac{n}{2}}}{n!}
\left(H^\dag_i h_i^\dag \right)^n |G \rangle_i |0\rangle_i \,,
\\
& \rightarrow \sum_{n} \frac{p^{\frac{n}{2}}}{n!}
\left(H^\dag_i\right)^n \left(\sqrt{\eta} h_i^\dag + \sqrt{1-\eta}
h_l^\dag\right)^n |G \rangle_i |0\rangle_i |0\rangle_l \,,
\label{Eqn:EMELossyOutput}
\end{align}
where we have performed the beamsplitter transformation $h_i^\dag
\rightarrow \sqrt{\eta} h_i^\dag + \sqrt{1-\eta} h_l^\dag$ on the
second line, with $h_l$ denoting the mode reflected from the
beamsplitter (i.e. the lost light).

Ultimately, the protocol for creating the EME states requires post
selection of the case where only one detector click is observed (for
each round of the protocol) and so the relevant terms in the above
expression are those with only a single photon in the $i$ mode.
However, by expanding out \eqn{Eqn:EMELossyOutput} it is clear that,
in the presence of loss, there will be many such terms, each
corresponding to a different total number of ensemble excitations,
$\left(H^\dag_i\right)^n$. Thus, in the final post selected state,
we expect to see additional terms which correspond to \emph{excess}
excitations in the two ensembles. This can be shown explicitly by
considering the full network for the EME preparation protocol,
together with the $\eta$-beamsplitters, tracing out the lost modes
(i.e. those reflected from the $\eta$-beamsplitters) and projecting
onto the case where only two detector clicks are observed. The
result is a state of the form
\begin{equation}
\rho_{EME,i,j} =  \left( \mathcal{E}_{H_i} \circ \mathcal{E}_{V_i}
\circ \mathcal{E}_{H_j} \circ \mathcal{E}_{V_j}\right)\left(
|EME\rangle_{i,j}\langle EME|\right) \,,
\label{Eqn:EMEExcitationErrors}
\end{equation}
where the excitation superoperators $\mathcal{E}_{S}$ are given, up
to an overall normalization factor, by
\begin{equation}
\mathcal{E}_{S} (\rho) = \sum_{i\ge 0} \frac{p^i(1-\eta)^i}{i!}
\left(S^\dag\right)^i \rho
\left(S^{}\right)^i\,,\label{Eqn:EMEExcitationSuperoperator}
\end{equation}
with $S^\dag$ the excitation operator for the corresponding ensemble
mode.


These excitation errors take the ensembles out of the logical qubit
basis and therefore will lead to errors in the computation.
Furthermore, it is not obvious that they can be detected with
certainty in subsequent `readout' stages of the protocol, since by
assumption the readout and detection efficiencies are less than
unity. However, by inspecting Eqs.
(\ref{Eqn:EMEExcitationErrors}) and (\ref{Eqn:EMEExcitationSuperoperator})
it is clear that the probabilities of the error terms are order
$p(1-\eta)$ or higher. Thus the relative contribution of these terms
can, in principle, be made arbitrarily small simply by reducing the
strength of the excitation laser pulse (which controls the parameter
$p$). This comes at the expense of decreasing the success
probability for preparing EME states, which incurs a linear increase
in the time required to prepare these states. Since EME preparation
occurs at the lowest level of the protocol, and since (given
sufficient physical resources) many EME states can be prepared in
parallel, these errors can be strongly suppressed without affecting
the overall efficiency of the computation. For the remainder of this
section we therefore treat the EME states as being essentially
perfect. In reality, of course, there will be some residual multiple
excitation errors, but provided the magnitude of these is
sufficiently small they can be dealt with within a more general
fault tolerance framework.

\subsection{Effect of loss in GHZ state preparation}

We now consider the effect of losses in the GHZ preparation step.
The inputs to this network are three two-qubit EME states, which we
assume to be perfect EME states, $\ket{\eme}_{i,j}=(H_i^\dag+V_j^\dag)(V_i^\dag+H_j^\dag)\ket{G}$. Our aim here is to demonstrate that, conditional
on observing three detector clicks in each of the (polarisation
resolving) detectors shown in \fig{EMEGHZ}, the remaining three
ensembles are projected into an ID-loss state. We make use of
similarities between our network for converting EME states into GHZ
states, and the circuit presented by Varnava et al. \cite{bib:Varnava07}
for generating GHZ states of photonic qubits.

As in the previous section, we model the photodetector and
ensemble-photon coupling efficiencies, $\eta_D$ and $\eta_E$, by
beamsplitters with the corresponding transmissivities placed at the
inputs to the detectors and outputs to the ensembles, as shown in
 \fig{fig:EMELoss}a (using the linear optical network enclosed by the dashed line in \fig{EMEGHZ}). The first step of the analysis
is to make use of the commutation relations for identical
beamsplitters to move the $\eta_E$ beamsplitters forwards through
the polarization rotators, and the $\eta_D$ beamsplitters backwards
through the network, to arrive at the equivalent network in
  \fig{fig:EMELoss}b, where the losses are now represented
by the three beamsplitters of effective transmissivity
$\eta=\eta_E\eta_D$.

\begin{figure}
\begin{center}
\includegraphics[width=0.8\columnwidth]{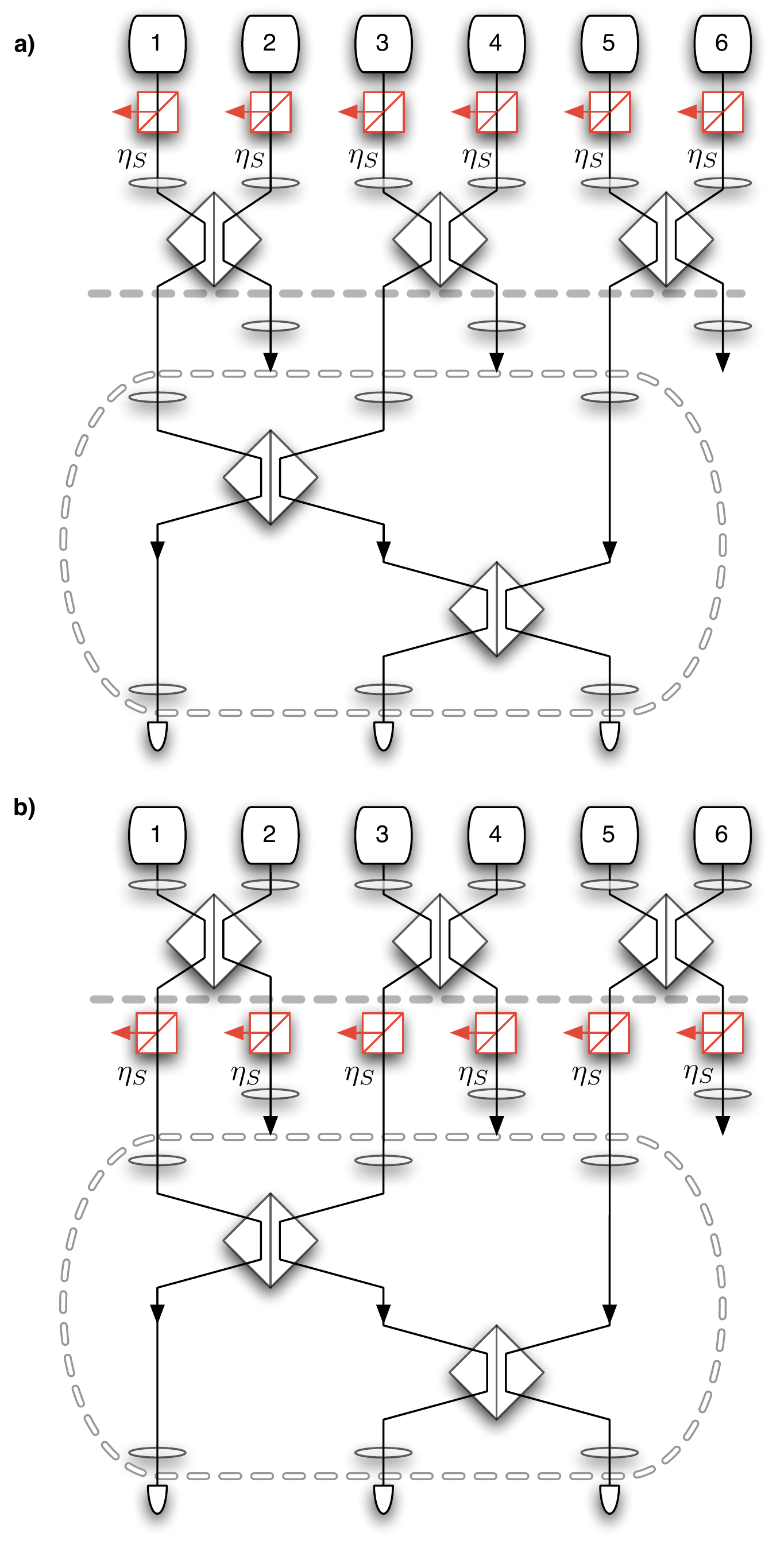
} \caption{The equivalence of independent losses (a) before and (b)
after the first row of PBS's in the GHZ preparation circuit of
Varnava \emph{et al}.  Single photon sources 1 to 6 emit an
$H$-polarised photon.  Immediately after the first row of PBSs in
(b) (at the dark dashed line) the photons are in the state
$\ket{\textrm{EME}}_{1,2}
\ket{\textrm{EME}}_{3,4}\ket{\textrm{EME}}_{5,6}$. }
\label{fig:VBRGHZcircuit}
\end{center}
\end{figure}

Varnava et al. showed, via a detailed analysis, that the all-optical
circuit of \fig{fig:VBRGHZcircuit}a, with imperfect single photon
sources of efficiency $\eta_S$ (which is unity for a perfect
source), leads to a state that is locally equivalent to a 3-qubit
ID-GHZ state on the qubits in modes 2, 3 and 5. This state is
obtained conditional on a single click being observed at each of the
detectors. Assuming that the detectors are perfect, Varnava et al.
showed that the output state is an ID-GHZ state with local loss rate
$f = 1-\frac{\eta_S}{2-\eta_S}$ acting on each qubit.


To make use of this result, first note that the circuit considered
by Varnava et al. is equivalent to the one shown in
\fig{fig:VBRGHZcircuit}b, in which the beamsplitters representing
source inefficiency are commuted through the first row of linear
optical elements. Now, if we consider the state of the six photons
at the position of the broken line in
\fig{fig:VBRGHZcircuit}b, it is found that they have the same form
as the three two-qubit EME states at the input to our GHZ network,
up to an unimportant polarization rotation acting independently on
each mode. Furthermore, the remainder of the optical network lying
below the broken line in \fig{fig:VBRGHZcircuit}b is
identical to the optical network used in our GHZ network, \fig{EMEGHZ}. Thus we
can directly apply the result of \cite{bib:Varnava07}, which implies that,
conditional on a single click being observed in each detector in our
network, the final state of ensembles 2, 4, and 6 is the ID-GHZ
state
\begin{multline}
\rho_{GHZ, 2,4,6} = \left( \mathcal{L}_{H_2} \circ \mathcal{L}_{V_2}
\circ \mathcal{L}_{H_4}  \circ \mathcal{L}_{V_4} \circ
\mathcal{L}_{H_6} \circ \mathcal{L}_{V_6}  \right) \ldots \\ \left(
|GHZ\rangle_{2,4,6}\langle GHZ|\right) \,.
\label{Eqn:GHZExcitationErrors}
\end{multline}
where the loss superoperators $\mathcal{L}_{S}$ are
\begin{equation}
\mathcal{L}_{S} (\rho) = (1-r) \rho + r (S \rho S^\dag +  S S^\dag
\rho S S^\dag) \,.\label{Eqn:GHZExcitationSuperoperator}
\end{equation}
The loss rate is given by $r=1-({2-\eta_E\eta_D})^{-1}$. Note this
is slightly different from the value $f$ determined by Varnava et
al., owing to the fact that we have not (as yet) included the effect
of coupling and detector efficiencies for the (as yet) unmeasured
ensembles 2, 4 and 6.

\subsection{Effect of loss in cluster state preparation and
measurement based computation}

To complete the demonstration of loss tolerance in our scheme, we
now consider the effect of loss when building large cluster states
with our destructive CZ gate, and in the single qubit measurement
phase of the computation.

First consider the effect of loss when attempting to fuse two
ID-cluster states via our destructive CZ gate, as shown in Fig.
\ref{Fusion}. As with the other gates, it is possible to model the
coupling and detector efficiencies with beamsplitters, and again the
$\eta_D$ beamsplitters can be commuted back through the optical
network such that both losses can be represented by a single
beamsplitter with transmisivity $\eta = \eta_E\eta_D$, at the output
of each ensemble. The CZ gate is non-deterministic, with success
heralded by the observation of two detector clicks in separate
detectors. Thus, provided the two input states are ID-cluster
states, the principal effect of the losses is to reduce the overall
success rate of the gate by a factor
$\eta/({2-\eta})$. Note that this rate is a product
of the loss rate, $r$, of the input ID-cluster states, together with
the additional losses incurred in the CZ gate itself. Upon success,
the resulting larger cluster state will be an ID state with the same
loss rate as the input states. This means that the CZ gate can be
used to build up large ID-cluster states of arbitrary shape, and,
assuming that additional losses incurred while these cluster states
are being constructed are negligible, the effective loss rate for
these clusters will be $r=1-({2-\eta})^{-1}$.

The other effect of loss in the CZ gate is that new `failure
outcomes' of the gate should now be considered. As well as the
original failure outcome - observing two photons in the same mode -
we must also consider the case when only one or zero clicks are
observed. In this case, the input state $|\Phi'\rangle$ in Fig.
\ref{Fusion} is left in an indeterminate (i.e. mixed) state.
However, it is generally possible to recycle significant parts of
the input clusters by performing a successful $Z$ measurement on
qubits `3' and `4'.

The final stage of our scheme where we must consider the effect of
loss is the `readout' phase of the measurement based computation. In
this phase, each qubit in the constructed cluster undergoes a single
qubit measurement, which comprises a `readout' pulse on the
ensemble(s), followed by some linear optical operations on the
output photons, and finally photodetection of the resulting photon
state. Clearly, losses will be important here since failure to
register a photon in one of these measurements will render the
remaining part of the cluster in an indeterminate state. Again, we
can make use of the results of Varnava et al. \cite{bib:Varnava07}, who
gave an explicit protocol for dealing with this loss model. This
involves encoding each logical qubit in a `tree structure'
comprising several physical qubits. Provided the total effective
loss rate for the computation (i.e. the combination of the
underlying loss rate of the initial cluster, combined with the
additional losses in the single qubit readout phase) is less than
$1/2$, efficient quantum computation is possible. This leads to the
final result: losses due to imperfect ensemble-photon coupling and
detector efficiencies can be tolerated provided
\begin{equation}
\frac{\eta}{{2-\eta}} > \frac{1}{2} \,,
\end{equation}
i.e. provided $\eta > 2/3$.


\section{Supplementary material: resource usage}

In order to give an estimate of the resources required by our
scheme, we calculate the cost of preparation of linear clusters in
terms of the total number of elementary laser pulses required to
build a linear cluster of length $N$. Although linear clusters are
not sufficient for performing arbitrary computations, they can serve
as a resource state for building up larger clusters of arbitrary
shape, with moderate additional overhead.

There are several stages of the protocol to consider, which we
consider independently. At the lowest level of the protocol we
prepare EME states from two separable ensembles. EME preparation
consists of two rounds, where each round is independent and
heralded. On average this requires $2/p$ pulses before success.

Now consider the resource consumption of the protocol for making
3-qubit cluster states of Fig. \ref{EMEGHZ}. This network takes as
an input three 2-qubit EME states. It then successfully bonds these
EME's together into a 3-qubit cluster with probability $1/32$. Thus
the number of operations (i.e. laser excitation pulses) per 3-qubit
cluster state is $3 \times 32 \times 2/p = 192/p$.

Next, suppose our procedure for building up large linear clusters is
as follows: we have a `main' cluster, which we are attempting to
grow, and smaller ancillary clusters which we attempt to bond onto
the main cluster. Because our destructive CZ gate succeeds with
probability 50\%, it can be seen that the ancillary clusters must be
of size at least four qubits if the main clusters is to grow on
average following each bonding attempt. Thus, we now turn our
attention to building 4-qubit ancillary clusters.

Preparing a 4-qubit linear cluster requires two 3-qubit linear
clusters. Given two such states they can be successfully bonded
together to form a 4-cluster with probability 50\%. Thus, the total
number of operations per 4-cluster is $2 \times 2 \times 192/p =
768/p$.

Given a resource of ancillary 4-clusters, the `main' cluster can be
grown at an average rate of 1/2 qubit per bonding attempt. So the
total number of operations per final qubit in the cluster is $2
\times N \times 768 / p = 1536 N / p$. Note that, given sufficient
experimental resources, many of the above steps can be performed in
parallel and so the actual number of time steps required to grow
cluster states can be much less than the total number of elementary
laser operations.

Here we have adopted an `incremental' approach to building up long
clusters, and at several stages we have assumed that upon a gate
failure, the remaining fragments of the ancillary clusters are not
recycled. Note that several alternate strategies could be employed
when preparing large clusters
\cite{bib:RohdeBarrett07,bib:Kieling06,bib:Gross06,kieling:130501}, 
 which might be significantly more resource efficient than
the incremental approach considered here.


\section{Supplementary material: Single qubit operations}

Throughout the description of the scheme in the main section of the
paper, we assumed that local unitary mode transformations can be
performed on each ensemble, or, in the case of the dual rail
encoding, on each pair of ensembles corresponding to a single qubit.
Such operations may be difficult to implement in practice,
particularly in the case of dual rail encoding. Here we describe how
these operations can in fact be deferred until the atomic
excitations are mapped onto photonic states.

Such single qubit operations are necessary at several stages of the
protocol: (1) during the production of the EME states; (2) during
the production of the three-qubit cluster states; (3) after CZ
operations between nodes of a cluster state; (4) during the
`readout' phase of the measurement based computation, when each
qubit is subject to a measurement of the observable
$\sin(\theta_i)X_i + \cos(\theta_i)Y_i$.

A generic state of $n$ atomic ensembles may be written
\begin{equation}
|\psi\rangle = f(H_1^\dag,V_1^\dag,\ldots H_i^\dag,V_i^\dag,\ldots
H_n^\dag,V_n^\dag) |G\rangle^{\otimes n} \nonumber
\end{equation}
where $f(\ldots)$ is a function of the excitation operators acting
on each ensemble. A local unitary mode transformation on the modes of the atomic ensemble, $U_i^{(a)}$,
transforms this state as
\begin{align}
U_i^{(a)}|\psi\rangle &= f(H_1^\dag,V_1^\dag,\ldots
{H'}_i^\dag,{V'}_i^\dag,\ldots H_n^\dag,V_n^\dag) |G\rangle^{\otimes
n} \nonumber
\end{align}
%
where
\begin{equation}
\left(
\begin{array}{c}
{H'}_i^\dag \\
{V'}_i^\dag
\end{array}
\right)= \mathbf{U}_i \left(
\begin{array}{c}
{H}_i^\dag \\
{V}_i^\dag
\end{array}
\right)\nonumber
\end{equation}
with $\mathbf{ U}_i$ a $2\times2$ unitary matrix. Subsequently, applying the
readout operation $R_i$ to the ensembles representing the $i$'th
qubit transforms $H_i^\dagger\rightarrow h_i^\dagger$ and $V_i^\dagger\rightarrow v_i^\dagger$:
\begin{align}
R_i U_i^{(a)}|\psi\rangle\ket{\vac}_i &= f(H_1^\dag,V_1^\dag,\ldots
{h'}_i^\dag,{v'}_i^\dag,\ldots 
) |G\rangle^{\otimes
n} \ket{\vac}_i \nonumber
\end{align}
with the photon operators given by
\begin{equation}
\left(
\begin{array}{c}
{h'}_i^\dag \\
{v'}_i^\dag
\end{array}
\right)= \mathbf{U}_i \left(
\begin{array}{c}
{h}_i^\dag \\
{v}_i^\dag
\end{array}
\right)\nonumber\,.
\end{equation}
Inspecting the above expressions, it is clear that $R_i
U_i^{(a)}|\psi\rangle\ket{\vac}_i = U_i^{(o)}
R_i|\psi\rangle\ket{\vac}_i$, where $U_i^{(o)}$ is a unitary mode transformation on the $i$th optical mode with the same unitary mode transformation matrix $\mathbf{ U}_i$  as $U_i^{(a)}$. In other words, local unitary mode
transformations of atomic ensemble modes can be deferred until after the readout pulse, and performed on the optical modes instead. Such transformations are straightforward to implement with
linear optical elements. The particular sequence of such operations
that must be performed generally depends on the outcomes of earlier
measurements in the scheme, and so a modest amount of classical
processing and optical switching is also required.


\begin{thebibliography}{27}
\expandafter\ifx\csname natexlab\endcsname\relax\def\natexlab#1{#1}\fi
\expandafter\ifx\csname bibnamefont\endcsname\relax
  \def\bibnamefont#1{#1}\fi
\expandafter\ifx\csname bibfnamefont\endcsname\relax
  \def\bibfnamefont#1{#1}\fi
\expandafter\ifx\csname citenamefont\endcsname\relax
  \def\citenamefont#1{#1}\fi
\expandafter\ifx\csname url\endcsname\relax
  \def\url#1{\texttt{#1}}\fi
\expandafter\ifx\csname urlprefix\endcsname\relax\def\urlprefix{URL }\fi
\providecommand{\bibinfo}[2]{#2}
\providecommand{\eprint}[2][]{\url{#2}}

\bibitem[{\citenamefont{Duan et~al.}(2001)\citenamefont{Duan, Lukin, Cirac, and
  Zoller}}]{bib:Duan01}
\bibinfo{author}{\bibfnamefont{L.-M.} \bibnamefont{Duan}},
  \bibinfo{author}{\bibfnamefont{M.~D.} \bibnamefont{Lukin}},
  \bibinfo{author}{\bibfnamefont{J.~I.} \bibnamefont{Cirac}}, \bibnamefont{and}
  \bibinfo{author}{\bibfnamefont{P.}~\bibnamefont{Zoller}},
  \bibinfo{journal}{Nature} \textbf{\bibinfo{volume}{414}},
  \bibinfo{pages}{413} (\bibinfo{year}{2001}).

\bibitem[{\citenamefont{Eisaman et~al.}(2004)\citenamefont{Eisaman, Childress,
  Andr\'e, Massou, Zibrov, and Lukin}}]{PhysRevLett.93.233602}
\bibinfo{author}{\bibfnamefont{M.~D.} \bibnamefont{Eisaman}},
  \bibinfo{author}{\bibfnamefont{L.}~\bibnamefont{Childress}},
  \bibinfo{author}{\bibfnamefont{A.}~\bibnamefont{Andr\'e}},
  \bibinfo{author}{\bibfnamefont{F.}~\bibnamefont{Massou}},
  \bibinfo{author}{\bibfnamefont{A.~S.} \bibnamefont{Zibrov}},
  \bibnamefont{and} \bibinfo{author}{\bibfnamefont{M.~D.} \bibnamefont{Lukin}},
  \bibinfo{journal}{Phys. Rev. Lett.} \textbf{\bibinfo{volume}{93}},
  \bibinfo{pages}{233602} (\bibinfo{year}{2004}).

\bibitem[{\citenamefont{Bali\'{c} et~al.}(2005)\citenamefont{Bali\'{c}, Braje,
  Kolchin, Yin, and Harris}}]{balic:183601}
\bibinfo{author}{\bibfnamefont{V.}~\bibnamefont{Bali\'{c}}},
  \bibinfo{author}{\bibfnamefont{D.~A.} \bibnamefont{Braje}},
  \bibinfo{author}{\bibfnamefont{P.}~\bibnamefont{Kolchin}},
  \bibinfo{author}{\bibfnamefont{G.~Y.} \bibnamefont{Yin}}, \bibnamefont{and}
  \bibinfo{author}{\bibfnamefont{S.~E.} \bibnamefont{Harris}},
  \bibinfo{journal}{Phys.\ Rev.\ Lett.} \textbf{\bibinfo{volume}{94}},
  \bibinfo{eid}{183601} (\bibinfo{year}{2005}).

\bibitem[{\citenamefont{Lan et~al.}(2007)\citenamefont{Lan, Jenkins,
  Chaneli\`{e}re, Matsukevich, Campbell, Zhao, Kennedy, and
  Kuzmich}}]{lan:123602}
\bibinfo{author}{\bibfnamefont{S.-Y.} \bibnamefont{Lan}},
  \bibinfo{author}{\bibfnamefont{S.~D.} \bibnamefont{Jenkins}},
  \bibinfo{author}{\bibfnamefont{T.}~\bibnamefont{Chaneli\`{e}re}},
  \bibinfo{author}{\bibfnamefont{D.~N.} \bibnamefont{Matsukevich}},
  \bibinfo{author}{\bibfnamefont{C.~J.} \bibnamefont{Campbell}},
  \bibinfo{author}{\bibfnamefont{R.}~\bibnamefont{Zhao}},
  \bibinfo{author}{\bibfnamefont{T.~A.~B.} \bibnamefont{Kennedy}},
  \bibnamefont{and} \bibinfo{author}{\bibfnamefont{A.}~\bibnamefont{Kuzmich}},
  \bibinfo{journal}{Phys.\ Rev.\ Lett.} \textbf{\bibinfo{volume}{98}},
  \bibinfo{eid}{123602} (\bibinfo{year}{2007}).

\bibitem[{\citenamefont{Chen et~al.}(2008)\citenamefont{Chen, Chen, Yuan, Zhao,
  Chuu, Schmiedmayer, and Pan}}]{Chen08}
\bibinfo{author}{\bibfnamefont{Y.-A.} \bibnamefont{Chen}},
  \bibinfo{author}{\bibfnamefont{S.}~\bibnamefont{Chen}},
  \bibinfo{author}{\bibfnamefont{Z.-S.} \bibnamefont{Yuan}},
  \bibinfo{author}{\bibfnamefont{B.}~\bibnamefont{Zhao}},
  \bibinfo{author}{\bibfnamefont{C.-S.} \bibnamefont{Chuu}},
  \bibinfo{author}{\bibfnamefont{J.}~\bibnamefont{Schmiedmayer}},
  \bibnamefont{and} \bibinfo{author}{\bibfnamefont{J.-W.} \bibnamefont{Pan}},
  \bibinfo{journal}{Nature Physics} \textbf{\bibinfo{volume}{4}},
  \bibinfo{pages}{103} (\bibinfo{year}{2008}).

\bibitem[{\citenamefont{Choi et~al.}(2008)\citenamefont{Choi, Deng, Laurat, and
  Kimble}}]{Choi08}
\bibinfo{author}{\bibfnamefont{K.~S.} \bibnamefont{Choi}},
  \bibinfo{author}{\bibfnamefont{H.}~\bibnamefont{Deng}},
  \bibinfo{author}{\bibfnamefont{J.}~\bibnamefont{Laurat}}, \bibnamefont{and}
  \bibinfo{author}{\bibfnamefont{H.~J.} \bibnamefont{Kimble}},
  \bibinfo{journal}{Nature} \textbf{\bibinfo{volume}{452}}, \bibinfo{pages}{67}
  (\bibinfo{year}{2008}).

\bibitem[{\citenamefont{Simon et~al.}(2007)\citenamefont{Simon, Tanji,
  Thompson, and Vuletic}}]{Simon07}
\bibinfo{author}{\bibfnamefont{J.}~\bibnamefont{Simon}},
  \bibinfo{author}{\bibfnamefont{H.}~\bibnamefont{Tanji}},
  \bibinfo{author}{\bibfnamefont{J.~K.} \bibnamefont{Thompson}},
  \bibnamefont{and} \bibinfo{author}{\bibfnamefont{V.}~\bibnamefont{Vuletic}},
  \bibinfo{journal}{Phys.\ Rev.\ Lett.} \textbf{\bibinfo{volume}{98}},
  \bibinfo{pages}{183601} (\bibinfo{year}{2007}).

\bibitem[{\citenamefont{Barrett and Kok}(2005)}]{bib:BarrettKok05}
\bibinfo{author}{\bibfnamefont{S.~D.} \bibnamefont{Barrett}} \bibnamefont{and}
  \bibinfo{author}{\bibfnamefont{P.}~\bibnamefont{Kok}},
  \bibinfo{journal}{Phys. Rev. A} \textbf{\bibinfo{volume}{71}},
  \bibinfo{pages}{060310(R)} (\bibinfo{year}{2005}).

\bibitem[{\citenamefont{Lim et~al.}(2005{\natexlab{a}})\citenamefont{Lim,
  Beige, and Kwek}}]{bib:Lim05}
\bibinfo{author}{\bibfnamefont{Y.~L.} \bibnamefont{Lim}},
  \bibinfo{author}{\bibfnamefont{A.}~\bibnamefont{Beige}}, \bibnamefont{and}
  \bibinfo{author}{\bibfnamefont{L.~C.} \bibnamefont{Kwek}},
  \bibinfo{journal}{Phys. Rev. Lett.} \textbf{\bibinfo{volume}{95}},
  \bibinfo{pages}{030505} (\bibinfo{year}{2005}{\natexlab{a}}).

\bibitem[{\citenamefont{Moehring et~al.}(2007)\citenamefont{Moehring, Maunz,
  Olmschenk, Younge, Matsukevich, Duan, and Monroe}}]{Moehring07}
\bibinfo{author}{\bibfnamefont{D.~L.} \bibnamefont{Moehring}},
  \bibinfo{author}{\bibfnamefont{P.}~\bibnamefont{Maunz}},
  \bibinfo{author}{\bibfnamefont{S.}~\bibnamefont{Olmschenk}},
  \bibinfo{author}{\bibfnamefont{K.~C.} \bibnamefont{Younge}},
  \bibinfo{author}{\bibfnamefont{D.~N.} \bibnamefont{Matsukevich}},
  \bibinfo{author}{\bibfnamefont{L.~M.} \bibnamefont{Duan}}, \bibnamefont{and}
  \bibinfo{author}{\bibfnamefont{C.}~\bibnamefont{Monroe}},
  \bibinfo{journal}{Nature} \textbf{\bibinfo{volume}{449}}, \bibinfo{pages}{68}
  (\bibinfo{year}{2007}).

\bibitem[{\citenamefont{Duan et~al.}(2002)\citenamefont{Duan, Cirac, and
  Zoller}}]{duan2002tdt}
\bibinfo{author}{\bibfnamefont{L.}~\bibnamefont{Duan}},
  \bibinfo{author}{\bibfnamefont{J.}~\bibnamefont{Cirac}}, \bibnamefont{and}
  \bibinfo{author}{\bibfnamefont{P.}~\bibnamefont{Zoller}},
  \bibinfo{journal}{Phys.\ Rev.\ A} \textbf{\bibinfo{volume}{66}},
  \bibinfo{pages}{23818} (\bibinfo{year}{2002}).

\bibitem[{\citenamefont{Raussendorf and Briegel}(2001)}]{bib:Raussendorf01}
\bibinfo{author}{\bibfnamefont{R.}~\bibnamefont{Raussendorf}} \bibnamefont{and}
  \bibinfo{author}{\bibfnamefont{H.~J.} \bibnamefont{Briegel}},
  \bibinfo{journal}{Phys. Rev. Lett.} \textbf{\bibinfo{volume}{86}},
  \bibinfo{pages}{5188} (\bibinfo{year}{2001}).

\bibitem[{\citenamefont{Nielsen}(2004)}]{bib:Nielsen04}
\bibinfo{author}{\bibfnamefont{M.~A.} \bibnamefont{Nielsen}},
  \bibinfo{journal}{Phys. Rev. Lett.} \textbf{\bibinfo{volume}{93}},
  \bibinfo{pages}{040503} (\bibinfo{year}{2004}).

\bibitem[{\citenamefont{Browne and Rudolph}(2005)}]{bib:BrowneRudolph05}
\bibinfo{author}{\bibfnamefont{D.~E.} \bibnamefont{Browne}} \bibnamefont{and}
  \bibinfo{author}{\bibfnamefont{T.}~\bibnamefont{Rudolph}},
  \bibinfo{journal}{Phys. Rev. Lett.} \textbf{\bibinfo{volume}{95}},
  \bibinfo{pages}{010501} (\bibinfo{year}{2005}).

\bibitem[{\citenamefont{Lim et~al.}(2005{\natexlab{b}})\citenamefont{Lim,
  Barrett, Beige, Kok, and Kwek}}]{bib:LimBarrett05}
\bibinfo{author}{\bibfnamefont{Y.~L.} \bibnamefont{Lim}},
  \bibinfo{author}{\bibfnamefont{S.~D.} \bibnamefont{Barrett}},
  \bibinfo{author}{\bibfnamefont{A.}~\bibnamefont{Beige}},
  \bibinfo{author}{\bibfnamefont{P.}~\bibnamefont{Kok}}, \bibnamefont{and}
  \bibinfo{author}{\bibfnamefont{L.~C.} \bibnamefont{Kwek}},
  \bibinfo{journal}{Phys. Rev. A} \textbf{\bibinfo{volume}{73}},
  \bibinfo{pages}{012304} (\bibinfo{year}{2005}{\natexlab{b}}).

\bibitem[{\citenamefont{Rohde and Barrett}(2007)}]{bib:RohdeBarrett07}
\bibinfo{author}{\bibfnamefont{P.~P.} \bibnamefont{Rohde}} \bibnamefont{and}
  \bibinfo{author}{\bibfnamefont{S.~D.} \bibnamefont{Barrett}},
  \bibinfo{journal}{New J. of Phys.} \textbf{\bibinfo{volume}{9}},
  \bibinfo{pages}{198} (\bibinfo{year}{2007}).

\bibitem[{\citenamefont{Kieling et~al.}(2006)\citenamefont{Kieling, Gross, and
  Eisert}}]{bib:Kieling06}
\bibinfo{author}{\bibfnamefont{K.}~\bibnamefont{Kieling}},
  \bibinfo{author}{\bibfnamefont{D.}~\bibnamefont{Gross}}, \bibnamefont{and}
  \bibinfo{author}{\bibfnamefont{J.}~\bibnamefont{Eisert}},
  \bibinfo{journal}{J. Opt. Soc. Am. B} p. \bibinfo{pages}{184}
  (\bibinfo{year}{2006}).

\bibitem[{\citenamefont{Gross et~al.}(2006)\citenamefont{Gross, Kieling, and
  Eisert}}]{bib:Gross06}
\bibinfo{author}{\bibfnamefont{D.}~\bibnamefont{Gross}},
  \bibinfo{author}{\bibfnamefont{K.}~\bibnamefont{Kieling}}, \bibnamefont{and}
  \bibinfo{author}{\bibfnamefont{J.}~\bibnamefont{Eisert}},
  \bibinfo{journal}{Phys. Rev. A} \textbf{\bibinfo{volume}{74}},
  \bibinfo{pages}{042343} (\bibinfo{year}{2006}).

\bibitem[{\citenamefont{Kieling et~al.}(2007)\citenamefont{Kieling, Rudolph,
  and Eisert}}]{kieling:130501}
\bibinfo{author}{\bibfnamefont{K.}~\bibnamefont{Kieling}},
  \bibinfo{author}{\bibfnamefont{T.}~\bibnamefont{Rudolph}}, \bibnamefont{and}
  \bibinfo{author}{\bibfnamefont{J.}~\bibnamefont{Eisert}},
  \bibinfo{journal}{Phys.\ Rev.\ Lett.} \textbf{\bibinfo{volume}{99}},
  \bibinfo{eid}{130501} (\bibinfo{year}{2007}).

\bibitem[{\citenamefont{Duan}(2002)}]{bib:Duan02}
\bibinfo{author}{\bibfnamefont{L.-M.} \bibnamefont{Duan}},
  \bibinfo{journal}{Phys. Rev. Lett.} \textbf{\bibinfo{volume}{88}},
  \bibinfo{pages}{170402} (\bibinfo{year}{2002}).

\bibitem[{\citenamefont{Varnava et~al.}(2007)\citenamefont{Varnava, Browne, and
  Rudolph}}]{bib:Varnava07}
\bibinfo{author}{\bibfnamefont{M.}~\bibnamefont{Varnava}},
  \bibinfo{author}{\bibfnamefont{D.~E.} \bibnamefont{Browne}},
  \bibnamefont{and} \bibinfo{author}{\bibfnamefont{T.}~\bibnamefont{Rudolph}}
  (\bibinfo{year}{2007}).

\bibitem[{\citenamefont{Rohde and Ralph}(2006)}]{bib:RohdeRalph06b}
\bibinfo{author}{\bibfnamefont{P.~P.} \bibnamefont{Rohde}} \bibnamefont{and}
  \bibinfo{author}{\bibfnamefont{T.~C.} \bibnamefont{Ralph}},
  \bibinfo{journal}{J. Mod. Opt.} \textbf{\bibinfo{volume}{53}},
  \bibinfo{pages}{1589} (\bibinfo{year}{2006}).

\bibitem[{\citenamefont{Dawson et~al.}(2005)\citenamefont{Dawson, Haselgrove,
  and Nielsen}}]{bib:Dawson05}
\bibinfo{author}{\bibfnamefont{C.~M.} \bibnamefont{Dawson}},
  \bibinfo{author}{\bibfnamefont{H.~L.} \bibnamefont{Haselgrove}},
  \bibnamefont{and} \bibinfo{author}{\bibfnamefont{M.~A.}
  \bibnamefont{Nielsen}}, \bibinfo{journal}{Phys. Rev. Lett.}
  \textbf{\bibinfo{volume}{96}}, \bibinfo{pages}{020501}
  (\bibinfo{year}{2005}).

\bibitem[{\citenamefont{Dawson et~al.}(2006)\citenamefont{Dawson, Haselgrove,
  and Nielsen}}]{bib:Dawson06}
\bibinfo{author}{\bibfnamefont{C.~M.} \bibnamefont{Dawson}},
  \bibinfo{author}{\bibfnamefont{H.~L.} \bibnamefont{Haselgrove}},
  \bibnamefont{and} \bibinfo{author}{\bibfnamefont{M.~A.}
  \bibnamefont{Nielsen}}, \bibinfo{journal}{Phys. Rev. A}
  \textbf{\bibinfo{volume}{73}}, \bibinfo{pages}{052306}
  (\bibinfo{year}{2006}).

\bibitem[{\citenamefont{Raussendorf et~al.}(2006)\citenamefont{Raussendorf,
  Harrington, and Goyal}}]{raussendorf2006fto}
\bibinfo{author}{\bibfnamefont{R.}~\bibnamefont{Raussendorf}},
  \bibinfo{author}{\bibfnamefont{J.}~\bibnamefont{Harrington}},
  \bibnamefont{and} \bibinfo{author}{\bibfnamefont{K.}~\bibnamefont{Goyal}},
  \bibinfo{journal}{Annals of physics} \textbf{\bibinfo{volume}{321}},
  \bibinfo{pages}{2242} (\bibinfo{year}{2006}).

\bibitem[{\citenamefont{Greenberger et~al.}(1990)\citenamefont{Greenberger,
  Horne, Shimony, and Zeilinger}}]{greenberger1990bst}
\bibinfo{author}{\bibfnamefont{D.}~\bibnamefont{Greenberger}},
  \bibinfo{author}{\bibfnamefont{M.}~\bibnamefont{Horne}},
  \bibinfo{author}{\bibfnamefont{A.}~\bibnamefont{Shimony}}, \bibnamefont{and}
  \bibinfo{author}{\bibfnamefont{A.}~\bibnamefont{Zeilinger}},
  \bibinfo{journal}{American J.\ of Phys.} \textbf{\bibinfo{volume}{58}},
  \bibinfo{pages}{1131} (\bibinfo{year}{1990}).

\bibitem[{\citenamefont{Hillery et~al.}(1999)\citenamefont{Hillery,
  Bu\ifmmode~\check{z}\else \v{z}\fi{}ek, and Berthiaume}}]{PhysRevA.59.1829}
\bibinfo{author}{\bibfnamefont{M.}~\bibnamefont{Hillery}},
  \bibinfo{author}{\bibfnamefont{V.}~\bibnamefont{Bu\ifmmode~\check{z}\else
  \v{z}\fi{}ek}}, \bibnamefont{and}
  \bibinfo{author}{\bibfnamefont{A.}~\bibnamefont{Berthiaume}},
  \bibinfo{journal}{Phys. Rev. A} \textbf{\bibinfo{volume}{59}},
  \bibinfo{pages}{1829} (\bibinfo{year}{1999}).

\end{thebibliography}
\end{document}